\documentclass[conference, 10pt,final]{IEEEtran}
\usepackage{setspace}
\usepackage{amssymb,graphicx,enumerate,amsfonts,amsmath,color,cite}
\usepackage[english]{babel}
\usepackage{pslatex}
\newcommand{\beq}{\begin{equation*}}
\newcommand{\eeq}{\end{equation*}}
\newcommand{\bea}{\begin{eqnarray}}
\newcommand{\eea}{\end{eqnarray}}
\newcommand{\beal}{\begin{align*}}
\newcommand{\eeal}{\end{align*}}
\newcommand{\bei}{\begin{itemize}}
\newcommand{\eei}{\end{itemize}}
\newcommand{\denk}{\begin{equation*} \begin{aligned}}
\newcommand{\denke}{\end{aligned}\end{equation*}}

\def\bary{\begin{array}}
\def\eary{\end{array}}
\def\bmat{\left[ \begin{array}}
\def\emat{\end{array} \right]}
\def\bmatt{\left\{ \begin{array}}
\def\ematt{\end{array} \right.}

\begin{document}
\title{Power Control in Multi-Layer Cellular Networks}
\author{\IEEEauthorblockN{Kemal Davaslioglu, Ender Ayanoglu\\}
\IEEEauthorblockA{Center for Pervasive Communications and Computing\\
Department of Electrical Engineering and Computer Science,
University of California, Irvine}}\maketitle
\begin{abstract} We investigate the possible performance gains of power control
in multi-layer cellular systems where microcells and picocells are
distributed within macrocells. Although multi-layers in cellular
networks help increase system capacity and coverage, and can
reduce total energy consumption; they cause interference, reducing
the performance of the network. Therefore, downlink transmit power
levels of multi-layer hierarchical cellular networks need to be
controlled in order to fully exploit their benefits. In this work,
we present an analytical derivation to determine optimum power
levels for two-layer cellular networks and generalize our solution
to multi-layer cellular networks. We also simulate our results in
a typical multi-layer network setup and observe significant power
savings compared to single-layer cellular networks.
\end{abstract}

\section{Introduction}
Power savings in cellular networks are not only going to reduce
operational expenses of operators but in addition, they will
significantly impact global carbon dioxide emissions. It has been
estimated that $2\%$ of global energy consumption is through
information and communication technologies \cite{Smart}. By the
increase in the offered data rates in the new standards such as
Long Term Evolution (LTE) and Worldwide Interoperability for
Microwave Access (WiMAX), and expected increase in subscribed
users, it is predicted that the power consumption of
telecommunications and mobile communications in particular, will
triple in the next decade \cite{Unplugged,Ericsson}. Approximately
$60$-$80\%$ of the total power consumed in mobile communications
is dissipated in the base stations, mainly through radio-frequency
(RF) conversion, power amplification and site cooling processes
\cite{Ericsson,Docomo}.

Application of multi-layer hierarchical cellular networks is
proposed in upcoming standards such as LTE and LTE-Advanced to
overcome the increasing demand in data rate and power
\cite{ReleaseNine,ReleaseTen}. Several low power base stations
such as microcells and picocells can be distributed within a
larger high power base station cell, namely the macrocell. Hence,
significant power savings are possible, high traffic loads can be
passed onto lower layers where high-rate low-power transmission is
possible, and any cell coverage problems such as shadowing effects
caused by buildings can be resolved. Also, depending on the
mobility pattern of users, coverage and excess handover traffic
issues can be eliminated. In addition, higher data rates can be
provided with careful network planning.

In this article, we investigate the deployment of two-layer
hierarchical networks and consider high-power macrocells overlaid
with low-power microcell systems, providing service within the
same cell. We present an analytical solution to determine optimum
power levels in these two-layer cellular systems. In cases where
channel conditions between microcells and users are superior
compared to macrocell base stations and users, macrocell base
stations can switch to a sleep mode and let microcells transmit
and receive data. This will provide significant power savings in
total energy consumption of the system, and at the same time,
deliver the same data rates to the users. We also generalize this
analytical solution to determine the optimum power levels in
multi-layer systems such as the one where macrocells, microcells
and picocells coexist. For cases where the optimal solution
exceeds permissible maximum power levels, we use another approach
proposed by Raman \emph{et al.} that uses linear programming (LP)
to determine the best solution given maximum power level
constraints of each base station \cite{RamanConf}. We also show
achievable power savings when multi-layers are deployed in the
system through a simple simulation setup.

In Section \ref{SystemModel}, the system models for single-layer
and two-layer cellular networks are introduced and in Section
\ref{PowerControlSection}, we present our analytical solution to
determine optimal power levels of both systems and generalize our
solution to multi-layer cellular systems. We also investigate the
necessary conditions for the existence of the proposed solutions.
In Section \ref{LPTitle}, we include maximum power levels of each
layer to the problem, and using these constraints, we explain how
to update base station power levels using the LP approach in
\cite{RamanConf}. We explain our simulation setup and its
parameters in Section \ref{Simulations} and show the power savings
for a two-layer architecture when compared to a single-layer
system. We conclude the paper with comments on the possible gains
that can be achieved with the deployment of multiple layers in
cellular networks.

\section{System Model}\label{SystemModel}
\begin{figure}[htb]\begin{center}
  \includegraphics[width=3.45in]{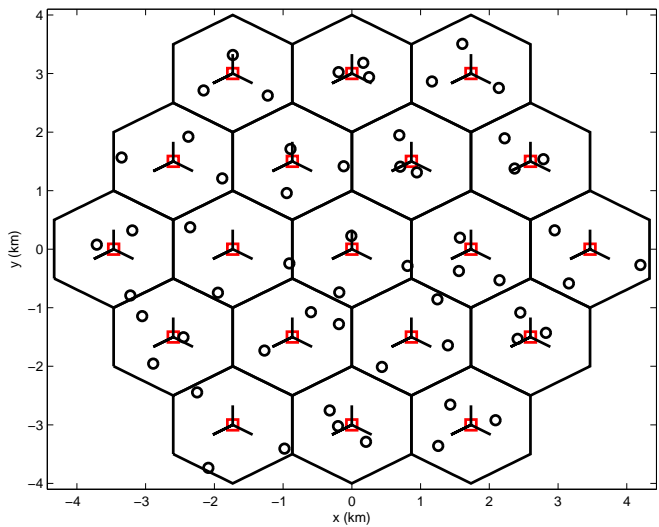}\\
  \caption{Single-layer network layout with $19$ hexagonal cells.
Every cell has three-sector $120^o$  directional antennas
positioned at the center of the cell. Each sector has one randomly
placed user. Squares and circles depict macrocell base stations
and users, respectively.}\label{baseline_sector}\end{center}
\end{figure}

In this section, we present system models for both single-layer
and two-layer cellular networks. A single-layer network
constitutes a reference scenario where we compare the gains with
respect to two-layer networks. To that end, we follow a simulation
setup similar to \cite{RamanConf} and consider a $19$-cell
hexagonal layout seen in Figure \ref{baseline_sector}. In each
cell, we locate a base station tower at the center of the hexagon
and deploy three sector antennas, each covering $120^o$ within the
cell. To mitigate the edge effects, the wrap-around technique is
employed \cite{ReleaseNine}. Furthermore, we assume that all base
stations in the system share the same resource that can be either
the same time slot, same frequency channel, same spreading code or
same time-frequency resource block as in LTE.

Users in the system are placed randomly within the cells
one-by-one such that there is only one user in each sector as in
\cite{RamanConf}. The condition that we emphasize is that the
generated user within each sector has to have the highest received
signal strength from its associated base station. If this
condition is not satisfied and the user needs to be handed over to
the neighboring base station, we discard the user and generate
another one. In total, we consider $57$ base stations and $57$
users in the system.

The first transmission is the learning phase for the channel
conditions in the system. All base stations send the same
predefined power level which we assumed as $5$ W. Assuming perfect
feedback, the central processor obtains all the channel gains
affecting every user. Then, the central processor discards a
predetermined number of users and considers them in outage so that
a common rate can be provided to the remaining users. The user
discarding procedure is based on the path loss criteria and the
users with the worst channel conditions are discarded. In the
power control step, the optimum power levels are calculated using
the analytical solution presented in the next section. The total
transmit power in the system after the power control step forms
the baseline system. We compare this value to the total transmit
power in the two-layer cellular system after power control step
for a fair comparison.

\begin{figure}[tb!]\begin{center}
  \includegraphics[width=3.5in]{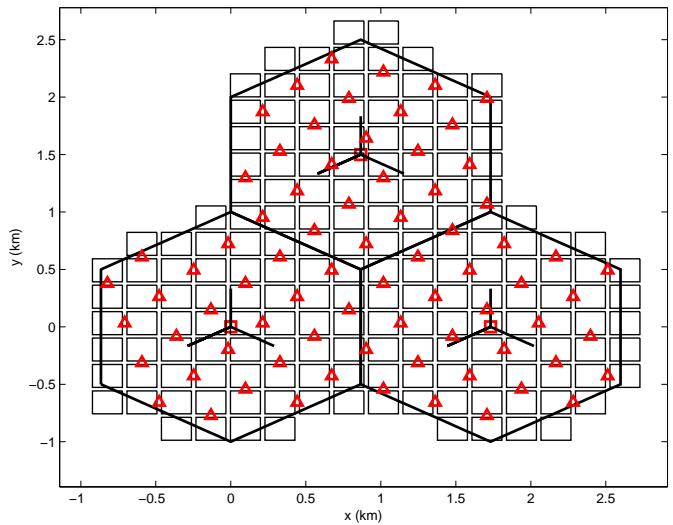}\\
  \caption{Two-layer hierarchical network layout with $19$ hexagonal cells overlaid with $467$ microcell base
    stations to model urban areas. Every cell includes 3-sector
    macrocell base stations placed in the center and depending on the geometry, $24-25$
    microcells with omnidirectional antennas are deployed in Manhattan grids. Squares and triangles depict
   macrocell and microcell base stations, respectively, and only $3$ cells are shown for clarity.}\label{twolayer_fig}\end{center}
\end{figure}

The two-layer cellular network includes additional microcell
overlay on top of the macrocell structure included in the baseline
system. We follow the simulation setup described in
\cite{Umts,ReleaseNine} for the microcell deployment in urban
areas and place microcell base stations on every other street on
the Manhattan grid where each block in the grid is $200$ m, each
street is $30$ m wide and microcells are deployed with
omnidirectional antennas. Figure \ref{twolayer_fig} depicts our
two-layer cellular system layout where high-power macrocells and
low-power microcells are deployed in the same area. User placement
methodology is the same as in single-layer networks that each user
is distributed randomly within each sector such that users have
the highest signal strength within the cell they are associated
with.
The first transmission is devoted to a learning phase for the
channel conditions in the system as it was for the single-layer
network. Using the feedback from every user, the central processor
forms the channel matrix of the system for both layers in the
system. Here, note that the channel matrix is an augmented channel
matrix which now includes the channel gain and path loss values
for both macrocell and microcell layers. Based on the channel
matrix, base station power levels for the two-layer cellular
system are updated. For those cases where the channel conditions
between users and microcell base stations are better than the
macrocell base stations, those macrocell base stations do not
transmit and switch to micro a sleep mode and let microcell base
stations transmit. In the sequel, this total transmit power value
after the power control step, including the sum of both layers is
compared with the baseline system to quantify the gains of
microcell deployment.

In the next section, we describe the analytical framework for the
power control process in single-layer, two-layer and multi-layer
cellular networks and comment on the necessary conditions for the
existence of the solutions.
\section{Power Control}\label{PowerControlSection}

\subsection{Single-Layer System}
In the single layer system, we only consider macrocells in the
system and we are interested in power savings in the downlink
since most of the power is dissipated in the base stations. The
total power transmitted in this single-layer system constitutes
our baseline case for comparison. We start our derivations by
identifying the signal-to-interference-plus-noise-ratio (SINR) for
user $i$ and write it as \bea \gamma_i =
\frac{g_{ii}p_i}{\sum\limits_{j \neq i} g_{ij}p_j + \sigma_i^2}
\eea where $p_i$ denotes the transmit power of $i$th base station
and $g_{ij}$ includes the path loss and shadowing observed by user
$i$ when base station $j$ transmits, $\sigma_i^2$ denotes the
noise. When we rearrange the terms above and divide by $g_{ii}$,
one would obtain \bea p_i = \gamma_i \sum\limits_{j \neq i}
\frac{g_{ij}}{g_{ii}}p_j + \gamma_i \frac{\sigma_i^2}{g_{ii}}.
\eea In vector-matrix form, the above equation can be written to
include all users as follows \bea \textbf{p} = \textbf{F}
\textbf{p} + \textbf{u} \eea where $ u_i = \gamma_i \sigma_i^2 /
g_{ii}$ and
\begin{align*} \textbf{F}  &= \begin{cases}
\gamma_i g_{ij}/g_{ii} & \text{if } j \neq i \\ 0 & \text{if } j =
i
\end{cases} \end{align*} and we will refer to $\textbf{F}$ as the normalized channel gain matrix for single-layer systems since it includes
channel gains between every user and every base station. In
\textbf{F}, except for the diagonal entries, the entries are
normalized by the channel gain between the user and its associated
base station. We can express the optimal solution by the vector
$\textbf{p}^*$, that minimizes the total power in the single layer
system as \bea \textbf{p}^* = (\textbf{I}_N - \textbf{F})^{-1}
\textbf{u} \label{opt_single}\eea where $\textbf{I}_N$ is an
$N\times N$ identity matrix. There are several comments to make
about this solution. First, the normalized channel gain matrix is
a nonnegative matrix since all its entries denote the channel gain
values and these are always non-negative. Second, we assume each
of these entries as realizations of the underlying stochastic
processes, namely they are all random path loss variables, such
that each channel gain in the system is determined independently
and hence, $\textbf{F}$ becomes a full-rank matrix. An important
remark is that one seeks a nonnegativity constraint on the
downlink transmit power levels of the base stations. Since
$\textbf{u}$ is always nonnegative, one needs to question the
existence and the nonnegativity of $(\textbf{I}_N -
\textbf{F})^{-1}$ for a feasible power level vector solution,
$\textbf{p}^*$.

In order to determine the existence and the nonnegativity of the
vector $\textbf{p}^*$, we apply the Perron-Frobenious theorem
\cite{TheoryOfMatrices}. This theorem seeks the irreducibility of
a matrix, therefore, one needs to check if $\textbf{F}$ is an
irreducible matrix or not. Following the same argument in
\cite{Goodman}, one sees that $\textbf{F}$ would be reducible if
and only if there exists more than one $0$ element on one row.
Since we include the channel gains of every base station to every
user using the wrap-around technique, we can conclude that
$\textbf{F}$ is irreducible. One direct result of the
Perron-Frobenious theorem is that for an irreducible nonnegative
matrix $\textbf{F}$, there always exists a positive real
eigenvalue $\lambda^*$ of $\textbf{F}$ and its associated
eigenvector where $\lambda^* = \max \{\lambda\}_{i=1}^N =
\rho(\textbf{F})$, is namely the spectral radius of $\textbf{F}$,
and every component of its associated eigenvector is nonnegative
\cite{TheoryOfMatrices}. Keeping these results in mind, one can
rewrite (\ref{opt_single}) such that
\begin{align} \textbf{p}^* = (\textbf{I}_N - \textbf{F})^{-1} \textbf{u} = \frac{1}{1-\rho(\textbf{F})} \textbf{u} \geq \textbf{0}\end{align}
and for the convergence of this solution we seek that the spectral
radius of $\textbf{F}$ needs to be less than 1,
$\rho(\textbf{F})<1$. Here, we note that in previous works lead by
Zander {et al.}, this condition was already identified in
\cite{ZanderCentral,Zender,Goodman,Foschini,GoodmanDPC,Bambos}.

A major drawback of this approach is that it is a centralized
solution. It is impractical for a central processor to have the
perfect knowledge of all path loss values for all users in the
system and determine the appropriate power levels for every user
and send back the power levels to the associated base stations in
a reasonable time. Therefore, several distributed solutions are
proposed where each base station can iterate and adjust its power
level using only the local acquired information without the need
for a global central processor \cite{Zender,Foschini,GoodmanDPC}.
One distributed power solution is proposed by Foschini and
Mijalcic where each base station updates its power level using the
following rule \cite{Foschini} \bea p_i^{n+1} =
\frac{\gamma_i}{g_{ii}} \left(\sum\limits_{j \neq i}g_{ij} p_j^n +
\sigma_i^2 \right) \eea where $p_i^n$ denotes the power level at
base station $i$ at $n$th iteration. In vector-matrix form, the
power update rule can be written as \bea \textbf{p}^{n+1} =
\textbf{F} \textbf{p}^{n} + \textbf{u}  \eea where
$\textbf{p}^{n}$ denotes the vector of macrocell transmit power
levels at $n$th iteration, and $\textbf{F}$ and $\textbf{u}$ are
as defined above. Note that, in \cite{Foschini} it has been shown
that for any initial power levels, using the above update rule,
base station power levels converge to the optimal solution after
several iterations.

\subsection{Two-Layer System}
For the two-layer system, we consider both macrocell and microcell
base stations and every user in the system experiences
interference from both layers. The SINR at user $i$ can be written
as \bea \gamma_i = \frac{g_{ii} p_i + h_{ii} q_i}{\sum_{j \neq i}
\left(g_{ij} p_j + h_{ij} q_{j} \right)+\sigma_i^2} \eea where
$p_i$ denotes the power transmitted from $i$th macrocell base
station and $q_i$ is the transmit power of microcell base station
$i$. The downlink channel coefficients for the path from macrocell
base station $j$ to user $i$ is denoted by $g_{ij}$ and for
microcell $j$ to user $i$ is shown by $h_{ij}$. The noise at
receiver $i$ is represented as $\sigma_i^2$. Using these, the
above equation can be further simplified as the following when we
rearrange terms and divide every term by $g_{ii}$
\[ p_i +
\frac{h_{ii}}{g_{ii}} q_i = \gamma_i \sum_{j \neq i}
\left(\frac{g_{ij}}{g_{ii}} p_j + \frac{h_{ij}}{g_{ii}} q_{j}
\right) + \gamma_i\frac{\sigma_i^2}{g_{ii}}. \] One can rewrite
the above equation in vector-matrix form as \begin{align}
\underbrace{\left[ \textbf{I}_N | \textbf{C}_{N\times N}
\right]}_{\textbf{A}_{N\times
2N}}\underbrace{\left[\bary{c}\textbf{p} \\ \textbf{q}
\eary\right]}_{\textbf{x}_{2N\times 1}} =
\underbrace{\left[\textbf{F}_{N\times N} | \textbf{G}_{N\times N}
\right]}_{\textbf{B}_{N\times 2N}}
\underbrace{\left[\bary{c}\textbf{p} \\ \textbf{q}
\eary\right]}_{\textbf{x}_{2N\times 1}}  +
\underbrace{\left[\bary{c} \gamma_1 \frac{\sigma_1^2}{g_{11}} \\ \vdots \\
\gamma_N \frac{\sigma_N^2}{g_{NN}} \eary\right]}_{\textbf{u}_{N
\times 1}} \end{align} where $N$ denotes the number of users in
the system, $\textbf{I}_N$ is an $N\times N$ identity matrix and
$\textbf{C}$ and $\textbf{G}$ matrices are as shown below
\begin{align} \textbf{C} &= \begin{cases} \frac{h_{ii}}{g_{ii}} & \text{if } j = i
\\ 0 & \text{if } j \neq i \end{cases}, \textbf{G}  = \begin{cases}
\gamma_i \frac{h_{ij}}{g_{ii}} & \text{if } j \neq i \\ 0 &
\text{if } j = i.
\end{cases}\end{align} We refer to $\textbf{H}$ as the normalized channel gain matrix for microcell
layer where the normalization is carried out with respect to
macrocell layer path loss values, $g_{ii}$.

For the analytical solution for the two-layer system, one seeks to
solve $\textbf{A} \textbf{x} = \textbf{B} \textbf{x} +
\textbf{u}$. Rearranging the terms on each side, the problem can
be restated as
\begin{align} \textbf{A} \left(\textbf{I}_{2N} - \widetilde{\textbf{B}} \right) \textbf{x} = \textbf{u} \end{align}
where $\textbf{B}  = \textbf{A}\widetilde{\textbf{B}}$ and
$\widetilde{\textbf{B}}$ is such that $\widetilde{\textbf{B}} =
\textbf{A}^{-1} \textbf{B}$. Also, $\textbf{A}^{-1}$ denotes the
adjoint matrix of the rectangular matrix $\textbf{A}$. Then, the
optimal solution for two-layer cellular system becomes
\begin{align} \label{opt_two} \textbf{x}^* = (\textbf{I}_{2N} - \widetilde{\textbf{B}})^{-1} \textbf{A}^{-1} \textbf{u}. 
\end{align}
Let us analyze the existence and nonnegativity of the optimal
solution vector, $\textbf{x}^*$. Following a similar analysis as
in the single-layer case, given that $\widetilde{\textbf{B}}$ is
nonnegative and irreducible, we can apply the Perron-Frobenious
theorem and see that there always exists a componentwise
nonnegative power vector $\textbf{x}^*$ given that the spectral
radius of $\widetilde{\textbf{B}}$ less than $1$. Since
$\textbf{A}$ is an $N \times 2N$ full-rank matrix, its adjoint
matrix $\textbf{A}^{-1}$ always exists. Therefore, we conclude
that as long as $\rho(\widetilde{\textbf{B}})<1$ condition is
satisfied, the above solution always yields nonnegative power
levels for two-layer cellular systems.

\subsection{Multi-Layer System}
In a multi-layer cellular network, we consider a system consisting
of macrocells, microcells and picocells. In this system, cross
layer interference becomes a serious issue and power levels of
each layer should be adjusted such that interference within the
same layer is minimized as well as the cross layer interference.
Following a similar analysis, we write the SINR at user $i$ that
includes interference from all layers \bea \gamma_i =
\frac{g_{ii}p_i + h_{ii}q_i+l_{ii} s_i}{\sum\limits_{j \neq i}
\left(g_{ij} p_j + h_{ij} q_j + l_{ij} s_j\right) +\sigma^2_i}\eea
where $l_{ij}$ denotes the channel gain from picocell transmitter
$j$ to user $i$, $s_j$ is the picocell transmit power and the rest
of the parameters are as defined as before. Rearranging the terms,
one obtains
\begin{align} p_i + \frac{h_{ii}}{g_{ii}} q_i + \frac{l_{ii}}{g_{ii}} s_i = \gamma_i \sum_{j \neq i} \left( \frac{g_{ij}}{g_{ii}} p_i + \frac{h_{ij}}{g_{ii}} q_{i} + \frac{l_{ij}}{g_{ii}}s_{i} \right) + \gamma_i\frac{\sigma_i^2}{g_{ii}} \end{align}
and this can also be written as
\begin{align} \underbrace{\left[ \textbf{I}_N | \textbf{C}_{N\times N} | \textbf{D}_{N\times N}\right]}_{\textbf{M}_{N\times 3N}}\underbrace{\left[\bary{c}\textbf{p} \\ \textbf{q} \\ \textbf{s}
\eary\right]}_{\textbf{y}_{3N\times 1}} =&
\underbrace{\left[\textbf{F}_{N\times N} | \textbf{G}_{N\times N}
| \textbf{H}_{N\times N}\right]}_{\textbf{N}_{N\times 3N}}
\underbrace{\left[\bary{c}\textbf{p} \\ \textbf{q} \\ \textbf{s}
\eary\right]}_{\textbf{y}_{3N\times 1}} \nonumber \\  & +
\textbf{u}_{N\times 1}
\end{align} where $\textbf{D}$ is a diagonal matrix with elements
$d_{ii} = l_{ii}/g_{ii}$ and \bea \textbf{L} &=
\begin{cases} \gamma_i \frac{l_{ij}}{g_{ii}} & \text{if } j \neq i
\\ 0 & \text{if } j = i \end{cases} \eea
and other terms are as defined previously. We refer $\textbf{H}$
as the normalized channel gain matrix for picocells where
macrocell path losses are used for normalization. Then, the
problem one needs to solve becomes $\textbf{M} \textbf{y} =
\textbf{N} \textbf{y} + \textbf{u}$ and the optimum power levels
in this multi-layer system would be \bea \textbf{y}^* =
(\textbf{I}_{3N}-\widetilde{\textbf{N}})^{-1} \textbf{M}^{-1}
\textbf{u} \label{opt_multi}\eea where $\widetilde{\textbf{N}}$ is
such that $\textbf{N} = \textbf{M} \widetilde{\textbf{N}}$ and
$\textbf{M}^{-1}$ is the adjoint matrix of $\textbf{M}$. For the
existence and nonnegativity conditions, we apply Perron-Frobenious
theorem and observe that \emph{iff}
$\rho(\widetilde{\textbf{N}})<1$ then, the above solution yields
feasible power levels.
\section{Power Level Constraints and LP Solution}\label{LPTitle}
In this section, we impose maximum power constraints due to
physical limitations arising at the low power base station radio
amplifiers. Every base station  amplifier has a certain peak power
level that depends on its specifications and one cannot exceed
this level. Depending on the number of layers in the cellular
system, we seek to solve (\ref{opt_single}), (\ref{opt_two}) or
(\ref{opt_multi}). In cases where the solution exceeds the maximum
transmit output power of base station, a different approach must
be pursued. Raman \emph{et al.} have proposed a solution to this
problem in \cite{RamanConf} where the power levels of two-layer
cellular system are adjusted using the linear programming approach
by imposing power constraints. We should note that in their
system, the two-layer cellular system consists of macrocells and
relays where relays are not connected to macrocells through
backhaul. In our case, we assume a backhaul connection between
macrocell and microcell base stations in the system. The problem
we seek to solve in two-layer hierarchical network system can be
formulated as
\begin{align}\label{lpsol}
     \begin{array}{rl}
     \min\limits_{\small{\begin{array}{c}p_1, \ldots, p_N \\ q_1,\ldots,q_N \end{array}}}  & \sum\limits_{i} \left( p_i + q_i
     \right)\\
     \mbox{\hspace{.1in} s.t. \hspace{.1in}} & \log_2 \left(1 + \frac{g_{ii} p_i + h_{ii} q_i}{\sum_{j \neq i} \left( g_{ij} p_j + h_{ij} q_{j}
     \right)+\sigma_i^2} \right) \geq r_i, \hspace{.05in} \forall i \\ & 0 \leq p_i \leq p_{\max}, \hspace{.05in} \forall i  \\ & 0 \leq q_i \leq q_{\max}, \hspace{.05in} \forall i
 \end{array}
\end{align}
where $r_i = \log_2(1+\gamma_i)$ bits/sec/Hz denotes the
transmission rate for user $i$, $p_{\max}$ and $q_{\max}$ are the
maximum transmit power levels of macrocells and microcells,
respectively. When the power levels in the two-layer network are
adjusted using the solution (\ref{lpsol}), one obtains the best
achievable performance given the maximum power level constraints
in each layer. This gives us an idea about how close we are to the
optimum solution in (\ref{opt_two}) where these power level
constraints did not exist.

\section{Simulations}\label{Simulations}
In this section, we first analyze the single-layer system and
compare the power gains with the two-layer cellular system. We
follow the simulation setup described in Section
\ref{SystemModel}, and without losing generality, consider a $19$
hexagonal cell layout seen in Figure \ref{baseline_sector}. In
each hexagonal-cell center, we place a base station tower with
three sector antennas, each covering $120^o$ within the cell and
the cell radius is assumed to be $1$ km as in \cite{ReleaseNine}.
Also, the wrap-around technique is employed to mitigate the edge
effects. We only consider horizontal radiation pattern for sector
antennas. The antenna gains are based on the angles between the
boresight direction of the base stations and mobile users and the
following antenna radiation pattern for the three-sector antenna
is used in our simulations
\begin{align} A(\theta) = - \min \left(12 \left( \frac{\theta}{\theta_{3\mathrm{dB}}}\right)^2, A_{\max}\right) \hspace{.05in} (\mathrm{dBi}) \end{align}
where $-180 \leq \theta \leq 180$, $\theta_{3\mathrm{dB}}$ and
$A_{\max}$ denote the $3$ dB beam width and the maximum
attenuation, respectively, and they are taken as
$\theta_{3\mathrm{dB}}=65^o$ and $A_{\max} = 20$ dB \cite{Umts}.

Users in the system are placed randomly within cells using the
procedure described in Section \ref{SystemModel} and we place one
user per sector making a total of $57$ users in the system sharing
the same resource. We target a common rate $r_0 = 1$ bps/Hz for
all users. In the first transmission, all base stations transmit
$5$ W and assuming a perfect feedback from every user, central
processor obtains the path loss values affecting every user. Based
on this information, we discard a fixed number of users with worst
SINR conditions. In our simulations, we investigate a wide range
of outage starting from $5\%$ to $17.5\%$ that corresponds to $3$
to $10$ discarded users out of $57$. For the remaining users, base
stations update their power levels using the analytical solution
in (\ref{opt_single}) such that both the interference and total
power levels are minimized. Hence, this total transmit power in
the power control step constitutes the baseline system and we
compare the gains of microcell deployments with respect to this
value.

The two-layer hierarchical system includes microcells on top of
the baseline system. Our simulation layout for two-layer cellular
systems is based on the path loss model for a microcell test
environment described in \cite{Umts} and we explain the details in
Section \ref{SystemModel}. The microcell layout is based on
Manhattan grids. As in the single-layer case, we simulate $57$
randomly distributed users, $57$ macrocell base stations and $467$
microcell based stations.
For the power control step, power levels of each base station are
updated using (\ref{opt_two}). For those cases, where the solution
in (\ref{opt_two}) yields exceeding power levels for microcells,
then LP solution using (\ref{lpsol}) is used such that solutions
within the permissible levels can be provided.

Due to different environment and terrain characteristics,
macrocell and microcell environments should be modeled distinctly.
Therefore, we consider different path loss models for macrocell
and microcell environments that have been accepted by European
Telecommunications Standards Institute (ETSI) and 3rd Generation
Partnership Project (3GPP) in \cite{ReleaseNine,Umts}. We refer
the reader to Annex B.1.8.1.2-3 in \cite{Umts} for details on
propagation model descriptions. We will omit explicit descriptions
and derivations due to space limitations and present the
propagation loss parameters used for both environments below.

The path loss model for macrocell users in urban areas is as
follows
\begin{align}\label{PLMacro} PL(\mathrm{dB})  = & \hspace{.05in} 40 \left(1-4\times
10^{-3}\Delta h_{bs}\right)\log_{10}(d) - 18\log_{10}(\Delta
_{bs}) \nonumber\\  & +  21\log_{10}(f) + 80
\end{align}
where $\Delta h_{bs}$ denotes the base station height measured
from average rooftop, $d$ denotes the difference between base
station and mobile user in km and $f$ is the carrier frequency in
MHz. In our simulations, the carrier frequency is taken as $2000$
MHz and all macrocell base station heights are assumed to be above
average rooftops and $\Delta h_{bs} = 15$ meters. The resulting
path loss formula for macrocell users can be expressed as
\begin{align} PL_{\mathrm{Macro}}(\mathrm{dB})  = 37.6 \log_{10}(d) + 128.15. \end{align}

Next, we present the path loss model for microcell users. We
assume that users are located outdoors and microcell base stations
are placed below rooftops. The following microcell propagation
model includes the effects of free space path loss that is denoted
by $PL_{\mathrm{FS}}$, diffraction effects from rooftops to
streets, $PL_{\mathrm{RTS}}$ and reductions caused by multiple
screen diffraction past rows of buildings, $PL_{\mathrm{MSD}}$ and
it is given by
\[ PL_{\mathrm{Micro}}(\mathrm{dB}) = PL_{\mathrm{FS}}(\mathrm{dB})
+ PL_{\mathrm{RTS}}(\mathrm{dB}) +
PL_{\mathrm{MSD}}(\mathrm{dB}).\]

Including the total effect of these three sources, the resulting
path loss equation reduces to
\begin{align}\label{MicroEqnOrg}
PL_{\mathrm{Micro}}(\mathrm{dB}) = 40 \log_{10} (d) + 30
\log_{10}(f) + 49 \end{align} where $d$ and $f$ are in km and MHz,
respectively, and in deriving the above equation, the following
assumptions are made. As will be explained shortly, base stations
are placed $5$ m below average rooftop, $\Delta h_{bs} = -5$ m,
mobile user and antenna height difference is $\Delta h_{ms} =
10.5$ m, horizontal distance between the mobile and the
diffracting edges is taken as $x = 15$ m, and building spacing is
$b = 80$ m. These parameter values are taken from \cite{Umts} and
they are commonly used in modelling microcell path loss in urban
and suburban environments. In our simulations, we have previously
assumed the carrier frequency to be $f = 2000$ MHz. Then,
(\ref{MicroEqnOrg}) reduces to
\begin{align}\label{MicroEqn}
PL_{\mathrm{Micro}}(\mathrm{dB}) = 40 \log_{10} (d) + 148.
\end{align}

Furthermore, we also include another important parameter in our
simulation model, the minimum coupling loss (MCL) which defines
the minimum possible propagation loss between base station and
mobile users. For macrocell and microcell environments, MCL values
are taken as $70$ dB and $53$ dB, respectively \cite{ReleaseNine}.
Then, the received power at mobile user can be written as
\begin{align} \mathrm{P_{RX}} = \mathrm{P_{TX}} - \max \left(PL- G_{\mathrm{TX}} - G_{\mathrm{RX}}, \mathrm{MCL} \right)\nonumber\end{align}
where $\mathrm{P_{TX}}$ and $\mathrm{P_{RX}}$ represent the
transmit power from base station regardless of its layer and
received power at mobile user, respectively. $PL$ denotes either
the microcell or macrocell path loss plus log-normal shadowing,
$G_{\mathrm{TX}}$ and $G_{\mathrm{RX}}$ are the transmit and
receive antenna gains, respectively, and they are taken as
$G_{\mathrm{TX}} = 11$ dB and $G_{\mathrm{RX}} = 0$ dB
\cite{ReleaseNine}. We consider maximum downlink transmit power
level of $43$ dBm and $33$ dBm for macrocell and microcell base
stations, respectively, and these values are in accordance with
the reference values in \cite{ReleaseNine}. We have considered
log-normal shadowing with standard deviation of $10$ dB for both
environments.
\begin{figure}[tb!]\begin{center}
  \includegraphics[height=2in,width=3.45in]{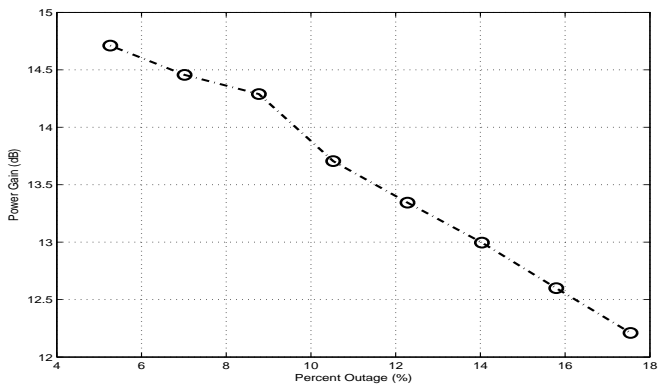}\\
  \caption{Two-layer cellular system power gains compared to a single layer are displayed versus percent outage for a $57$ user system.}\label{figresults}\end{center}
\end{figure}

Figure \ref{figresults} shows average power gains for various
outage percentages for $3$ to $10$ discarded users out of $57$
user system and this corresponds to $5\%$ to $17.5\%$ outages.
Over $1000$ simulations, we observed substantial average power
gains around $12-15$ dB compared to the baseline system. We also
note that as we discarded more users, average power gains
decreased due to decreasing degrees of freedom in the system. The
previous work reported by Raman \emph{et al.} shows average gains
around $3$ dB when relays are deployed to predefined locations in
a $19$-cell hexagonal layout \cite{RamanConf}. The reasons for the
difference between their results and in this paper is twofold.
First is that in our setup, microcell base stations are always
connected to macrocells through backhaul and microcells do not
have to wait for macrocell transmission to decode and forward the
message. This increases the degrees of freedom that the central
processor has since it has more variables to further optimize the
power levels. Second, there are more microcell base stations
considered in our simulation setup and this brings better channel
conditions between microcell base stations and users on average.

\section{Conclusion}\label{Conclusions}
Multi-layer hierarchical cellular networks offer significant
savings in total transmit power. This reduces operational expenses
for the operators and at the same time, decreases global carbon
dioxide emissions. Moreover, employing multiple layers in networks
also helps increase coverage within the cell as well as providing
higher transmission rates for low-mobility users and continuous
service for vehicular users. On the other hand, increasing layers
in the cellular systems creates detrimental interference within
the same layer as well as in the cross layers. In this paper, we
presented an analytical derivation to determine optimum power
levels for two-layer networks that would minimize the total
transmit power in the system and at the same time provide the same
data rates. We also extended this analysis to multi-layer systems
where macrocells, microcells and picocells are employed together.
Through simulations we showed that significant power gains are
possible when two-layers are employed in the cellular systems
instead of a single-layer.

\end{document}